\documentclass[twocolumn,aps,prl,amsmath,amssymb]{revtex4-1}
\usepackage{newtxtext}
\usepackage{newtxmath}
\usepackage{microtype}
\usepackage{graphicx}
\usepackage{dcolumn}
\usepackage{bm}
\usepackage{multirow}
\usepackage{amssymb}
\usepackage{amsmath}
\usepackage{xcolor}
\usepackage[colorlinks,citecolor=blue,linkcolor=red,urlcolor=blue]{hyperref}
\usepackage{braket}
\usepackage{CJK}
\usepackage{indentfirst}
\usepackage{cases}
\usepackage{mathrsfs}
\usepackage{orcidlink}

\def\Fig#1{Fig.~\ref{#1}}

\begin{document}
\title{Symmetric Mass Generation Transition and its Nonequilibrium Critical Dynamics in a Bilayer Honeycomb Lattice Model}
\author{Zhi-Xuan Li\,\orcidlink{0009-0007-5861-0257}$^{1,2}$}
\author{Yin-Kai Yu\,\orcidlink{0009-0006-2589-2772}$^{3,4}$}
\author{Zi-Xiang Li\,\orcidlink{0000-0002-3941-365X}$^{3,4}$}
\email{zixiangli@iphy.ac.cn}
\author{Shuai Yin\,\orcidlink{0000-0001-8534-9364}$^{1,2}$}
\email{yinsh6@mail.sysu.edu.cn}

\affiliation{$^{1}$School of Physics, Sun Yat-sen University, Guangzhou 510275, China}
\affiliation{$^{2}$Guangdong Provincial Key Laboratory of Magnetoelectric Physics and Devices, School of Physics, Sun Yat-sen University, Guangzhou 510275, China}
\affiliation{$^{3}$Beijing National Laboratory for Condensed Matter Physics \& Institute of Physics, Chinese Academy of Sciences, Beijing 100190, China}
\affiliation{$^{4}$University of Chinese Academy of Sciences, Beijing 100049, China}

\date{\today}
\begin{abstract}
Symmetric mass generation (SMG) transitions defy the conventional Landau-Ginzburg-Wilson paradigm by opening a many-body gap without spontaneous symmetry breaking or topological order,  attracting intense interest across particle physics and condensed matter physics. Here, we utilize unbiased quantum Monte Carlo simulations to investigate the equilibrium and nonequilibrium critical dynamics of the SMG transition in a bilayer honeycomb lattice model. We unambiguously confirm the existence of an SMG transition at $J_{\text{c}}=2.584(8)$ that separates the Dirac semimetal phase from a symmetry-preserving SMG phase. High-precision extraction of the critical exponents reveals a novel universality class that profoundly departs from mean-field theory. We then extend our study to the nonequilibrium regime, exploring the driven dynamics of the SMG transition. Notably, despite the breakdown of the prerequisites for the celebrated Kibble-Zurek mechanism, the nonequilibrium SMG transition still follows the generalized finite-time scaling. By bridging equilibrium criticality and nonequilibrium dynamics, our work uncovers the universal critical properties of SMG transitions, providing a solid theoretical basis for future experimental studies of SMG physics.
\end{abstract}

\maketitle

{\it Introduction}---The origin of mass ranks among the core issues in fundamental physics. Within the framework of traditional theories, spontaneous symmetry breaking plays a pivotal role: bosons gain mass through the well-known Higgs mechanism, whereas fermions acquire mass via the spontaneous symmetry breaking of bosonic fields whose nonzero expectation value imparts mass to fermions through Yukawa coupling~\cite{Weinberg1996Modern}. Symmetric mass generation (SMG) offers a novel mechanism for fermions to acquire mass without breaking any symmetries~\cite{wangsym2022}. Furthermore, the SMG phase transition represents a fermionic version of the deconfined quantum critical point~\cite{yzyprx2018,yzyprb2018}, wherein the dominant fluctuations are fractionalized degrees of freedom, transcending the conventional Landau-Ginzburg-Wilson (LGW) paradigm of phase transitions~\cite{Senthil2004a,Senthil2004b}. Accordingly, SMG and its critical behavior have garnered growing attention in recent years, particularly within the fields of particle physics and condensed matter physics~\cite{houprb2023,heprb2016,slagleprb2015,wxcarxiv2023,guo2026arxiv,ben2015,yzyprb2015,zengprl2022,ayyarprd2016,ayyarprd2017,yzyprx2018,yzyprb2018,houprb2023,heprb2016,wangprr2020,tong2022,wangsym2022,Wang2014science,xu2021arxiv,LU2023prb,Lu2023prb2,You2022PRL3450model,SMGReview2025,Nouman2025PRL,lu2023arxiv}.

Lattice models provide powerful platforms for investigating SMG transitions~\cite{houprb2023,slagleprb2015,heprb2016,wxcarxiv2023,guo2026arxiv}. However, despite extensive research, the existence and precise nature of these transitions remain controversial in microscopic models. In particular, based on the variational Monte Carlo (VMC) method, the bilayer honeycomb lattice model was predicted to host a SMG transition between the Dirac semimetal (DSM) phase and the symmetric insulating phase~\cite{houprb2023}. However, the VMC approach relies strongly on the choice of the variational trial state, which may introduce unquantified systematic biases into the resulting predictions. Furthermore, the universal critical exponents governing this putative SMG transition remain largely unresolved. Therefore, an unbiased and systematic investigation of this model is imperative to advance our fundamental understanding of SMG physics, potentially catalyzing the establishment of a generalized theory of quantum criticality beyond the LGW paradigm.

Meanwhile, nonequilibrium critical dynamics not only serves as a key research topic in fundamental statistical physics but also plays a vital role in the preparation and realization of novel quantum states of matter~\cite{Jacek2010aip,Polkovnikov2011rmp,Luca2016aip,Mitra2018arocmp}. A typical kind of nonequilibrium dynamics involves linearly tuning a parameter to cross the critical point. In conventional symmetry-breaking phase transitions, the celebrated Kibble-Zurek mechanism (KZM)~\cite{kibble1976,zurek1985} and its full scaling form generalization, known as finite-time scaling (FTS)~\cite{zhongprb2005,Gongnjp2010,fengprb2016,huangprb2014,yinprb2014}, have established a general mechanism for driven dynamics and sparked extensive research and generalization~\cite{Zoller2005prl,Dziarmaga2005prl,Polkovnikov2005prb,liuprb2014,liuprl2015,yinprl2017,lipre2019,Pickett1996,Ruutu1996,Monacoprl2002,Lee2024,Deng_2008,grandiprb2010,Chandranprb2012,Keesling2019,grandiprb2011,Shu2023,zzarxiv2024,zengFTSKZM2025,Du2023,Ko2019,Maegochi2022prl,Ebadi2021,Qiu2022sa,Ebadi2022science,Sunami2023science,Li2023PRX,Clark2016science,Huse2012prl,Dupont2022prb,Polkovnikov2008natphy,Deng2025prl,Wang2025cpl,Wang2025nc,joseprb2024,Zeng2023prl}. However, because the origin KZM and FTS are rooted in the paradigm of spontaneous symmetry breaking (SSB), their applicability to SMG transitions remains an open and important question. Therefore, given the fundamental significance of both SMG transitions and the KZM, an in-depth investigation into the nonequilibrium dynamics of SMG transitions is highly demanded.



\begin{figure}[tbp]
	\centering
	\includegraphics[width=\linewidth,clip]{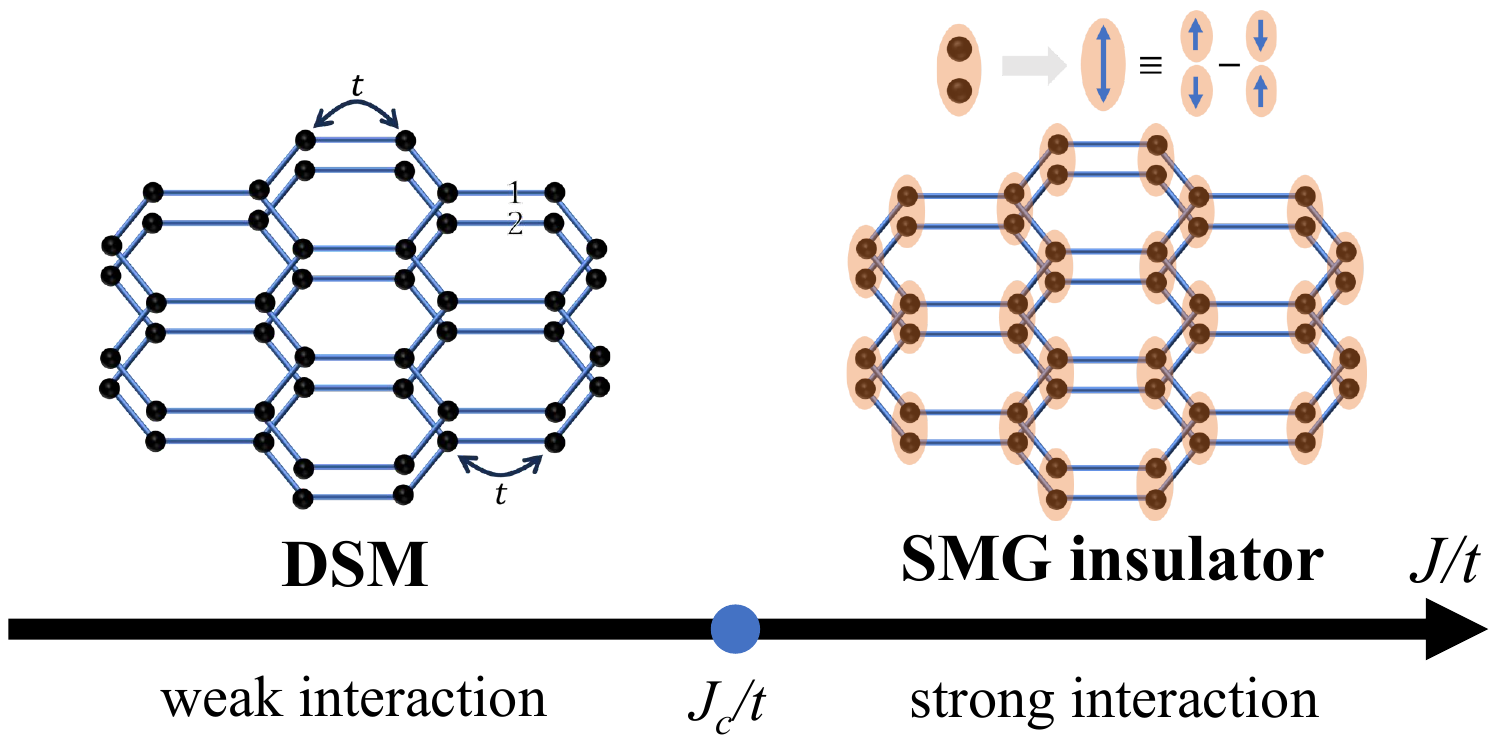}
	\vskip-3mm
	\caption{{\bf Sketch of the phase diagram.} The system undergoes a phase transition from DSM to SMG insulator phase at $J_{\text{c}}=2.584(8)$. In the weak-interaction regime ($J < J_{\text{c}}$), the system is in the DSM phase and hosts gapless Dirac fermions. In the strong-interaction regime ($J > J_{\text{c}}$), the fermions acquire an energy gap, driving the system into the SMG phase. The interlayer spin interaction causes interlayer spin singlets (orange ellipse), without breaking the symmetry of the system.}
	\label{figphase}
\end{figure}

In this Letter, we employ unbiased determinant quantum Monte Carlo (DQMC) simulations~\cite{blankprd1981,hirschprl1981,Assaad2008,Li2019QMC} to systematically study the bilayer honeycomb lattice model with interlayer spin coupling at half-filling. By performing comprehensive numerical calculations, we determine the ground-state phase diagram of the model and identify a phase transition between the Dirac semimetal (DSM) phase and a symmetric gapped phase. Crucially, we find that this transition proceeds without any signature of spontaneous symmetry breaking, as explicitly illustrated in~\Fig{figphase}, allowing us to unambiguously infer that the underlying driving mechanism is SMG. In addition, we determine the correlation length exponent $\nu$ and the anomalous dimension of the fermion $\eta$ as $\nu=0.945(5)$ and $\eta=0.11(2)$, respectively. We find that these exponents differ significantly from previous theoretical predictions, calling for further investigations of the corresponding analytical theories.

Beyond equilibrium studies, we further extend our investigation to the nonequilibrium regime by exploring the driven critical dynamics of the SMG transition. Note that the conventional KZM is rooted in the symmetry-breaking paradigm, wherein the key dynamical scale is set by the separation of topological defects at the intersections of different broken-symmetry domains~\cite{kibble1976,zurek1985}. Surprisingly, we find here that the FTS persists across the SMG phase transition, despite the complete absence of spontaneous symmetry breaking. On this basis, we generalize the KZM/FTS framework to nonequilibrium processes associated with SMG phase transitions for the first time. This extension not only validates the universality of KZM/FTS beyond conventional symmetry-breaking scenarios but also provides a novel theoretical framework for characterizing nonequilibrium SMG dynamics in lattice systems.

{\it Model}---We consider a bilayer honeycomb lattice model with antiferromagnetic interlayer spin interactions at half-filling. The Hamiltonian is expressed as follows~\cite{houprb2023}:
\begin{equation}
  H=-t\sum_{\langle ij \rangle \sigma l}(c^{\dagger}_{i\sigma l}c_{j\sigma l} +\text{h.c.})+J\sum_{i}\boldsymbol{S}_{i,1}\cdot\boldsymbol{S}_{i,2} ,
  \label{eq:H1}
\end{equation}
where $t$ and $J$ denote the hopping amplitude and interaction strength, respectively. $c^{\dagger}_{i\sigma l}$ $(c_{i\sigma l})$ represents the creation (annihilation) operator with spin polarization $\sigma = {\uparrow, \downarrow}$ and layer index $l = 1,2$. $\langle ij \rangle$ denotes a pair of nearest-neighbor sites, while $\boldsymbol{S}_{i,l}$ is defined as 
$\boldsymbol{S}_{i,l}=\frac{1}{2}c^{\dagger}_{i\alpha l}\boldsymbol{\sigma}^{\alpha\beta} c_{i\beta l}$ with $\boldsymbol{\sigma}=(\sigma_{x},\sigma_{y},\sigma_{z})$ being the Pauli matrices. In the following, $t$ is fixed to be unity. Crucially, this model is free from the fermion sign problem, enabling large-scale DQMC simulations to extract its ground-state properties in an unbiased manner~\cite{wuprb2005,Li2016PRL,wxcarxiv2023}.

\begin{figure*}[tbp]
	\centering
	\includegraphics[width=1.0\linewidth,clip]{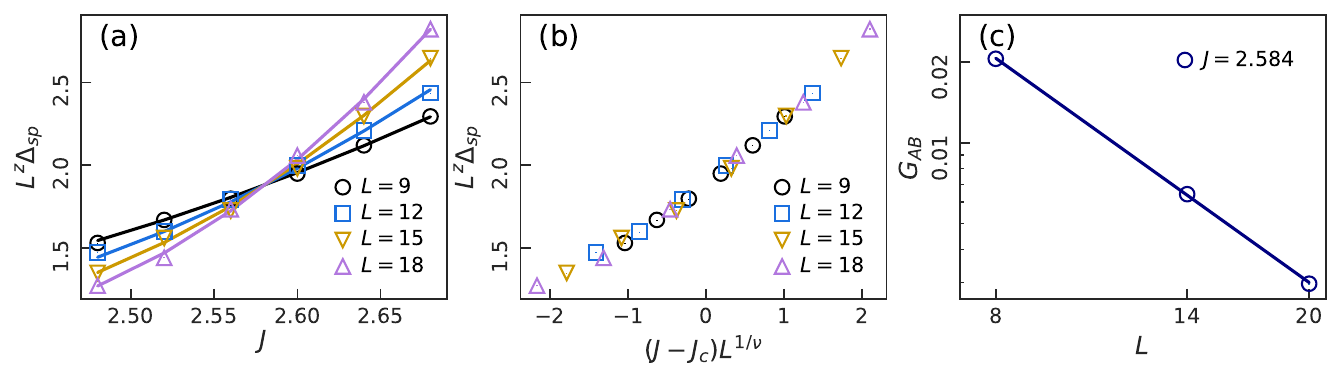}
	\vskip-3mm
	\caption{{\bf Determination of critical point and critical properties.} \textbf{(a)} The crossing point of $L^{z}\Delta_{\text{sp}}$ as a function of $J$ for various $L$ determines the critical point $J_{\text{c}}=2.584(8)$. \textbf{(b)} The correlation-length exponent $\nu$ is estimated to be $\nu = 0.945(5)$ from the scaling collapse of the rescaled curves of $L^{z}\Delta_{\text{sp}}$ versus $(J-J_{\text{c}})L^{1/\nu}$. \textbf{(c)} The fermion correlation function $G_{\text{AB}}$ at $J_{\text{c}}=2.584(8)$ is shown as a function of system size $L$ on a log-log scale. Power-fitting gives the anomalous dimension $\eta$ as $\eta=0.11(2)$.}
	\label{figure1}
\end{figure*}

As shown in \Fig{figphase}, considering different values of $J/t$, when $J/t\rightarrow 0$, the system is in the DSM phase, where gapless Dirac fermions are present. As $J/t \rightarrow\infty$, the gapless Dirac fermions acquire a gap, leading the system to an insulating state known as the SMG insulator, for which the ground state wave function is $|\psi \rangle = \prod_{i}(c^{\dagger}_{i1\uparrow}c^{\dagger}_{i2\downarrow}-c^{\dagger}_{i1\downarrow}c^{\dagger}_{i2\uparrow})|0 \rangle$.  Since this transition occurs without breaking any symmetries of the model, a direct transition between these two phases is, by definition, an SMG transition. However, although the VMC results suggest that this model undergoes an SMG transition, the VMC approach is inherently biased, as it relies strongly on the choice of the variational trial state. Consequently, the presence of an intervening symmetry-breaking phase cannot be definitively ruled out. Specifically, the primary candidate suggested by mean-field analysis is the interlayer exciton condensation (EC) order~\cite{houprb2023,lu2023arxiv}, whose order parameter is defined as $\phi_{\text{EC}}\equiv \sum_{i\sigma}(-)^{i}c^{\dagger}_{i\sigma 1}c_{i\sigma 2}$, where $i$ is even (odd) for sublattice A (B).

{\it Equilibrium criticality}---Firstly, we explore the ground state properties of this model (\ref{eq:H1}) via the DQMC method. In contrast to VMC used in Ref.~\cite{houprb2023}, the QMC approach employed here is unbiased. If a SMG phase transition exists, there is no corresponding order parameter to characterize it. Thus, the single-particle energy gap $\Delta_{\text{sp}}$ becomes an important indicator for characterizing this phase transition. In the thermodynamic limit, $\Delta_{\text{sp}} = 0$ indicates that the system is in the DSM phase, whereas $\Delta_{\text{sp}} > 0$ signifies a transition to the SMG insulator phase.

For finite-size systems, one can construct a dimensionless quantity as $L^z\Delta_{\text{sp}}$ with $z=1$ being the dynamic exponent to probe the critical point. According to the general scaling theory, near the critical point $J_{\text{c}}$, $L^z\Delta_{\text{sp}}$ should satisfy
\begin{equation}
L^{z}\Delta_{\text{sp}} = \mathcal{F}((J-J_{\text{c}})L^{1/\nu}) ,
	\label{eq2}
\end{equation}
in which $\mathcal{F}$ is the scaling function and $\nu$ is the correlation length exponent. When $J=J_{\text{c}}$, $L^{z}\Delta_{\text{sp}}$ is a constant independent of $L$, allowing us to precisely pinpoint the location of the critical point.

To extract $\Delta_{\text{sp}}$, we numerically compute the single-particle imaginary-time Green’s function $G(\boldsymbol{K},\tau)=\langle c_{\boldsymbol{K}}(\tau)c^{\dagger}_{\boldsymbol{K}}(0)\rangle$, in which $\boldsymbol{K}$ denotes the momentum of the Dirac point. For sufficiently large $\tau$, $G(\boldsymbol{K},\tau)$ satisfies $G(\boldsymbol{K},\tau)\propto e^{-\Delta _{\text{sp}}(\boldsymbol{K})\tau}$. Accordingly, $\Delta_{\text{sp}}$ can be obtained by performing an exponential fit to $G(\boldsymbol{K},\tau)$ (see details in the Supplementary Materials~\cite{SupMat}). The numerical results are shown in Fig.~$\ref{figure1}$~(a), in which curves of $L^{z}\Delta_{\text{sp}}$ versus $J$ for different $L$ almost intersect at a single crossing point, demonstrating that a phase transition indeed occurs at the critical point $J_{\text{c}}=2.584(8)$. 


We proceed to determine the critical exponents associated with this quantum phase transition. 
First, the correlation length exponent $\nu$ is obtained via a finite-size scaling analysis 
of the single-particle gap, $\Delta_{\text{sp}}$. By rescaling the data according to the ansatz 
$L^z \Delta_{\text{sp}} = f((J-J_{\text{c}})L^{1/\nu})$, we tune $\nu$ to achieve the best data collapse for 
different system sizes $L$. As illustrated in Fig.~\ref{figure1}~(b), this procedure yields 
$\nu = 0.945(5)$. Second, we determine the anomalous dimension $\eta$ by computing the fermion 
correlation function at the critical point, $J_{\text{c}}=2.584(8)$. This function, defined as 
$G_{\text{AB}}(L)\equiv L^{-2}\sum_{r'}\langle \hat{c}^{\dagger}_{\text{B},r'+r_{m}}\hat{c}_{\text{A},r'} 
+ \text{h.c.}\rangle$ for sublattices A and B at maximum separation $r_{m}=(L/2,L/2)$ 
\cite{sekiprb2019}, is expected to decay as a power law. A fit of our numerical data to 
the scaling relation $G_{\text{AB}}(L)\propto L^{-(2+\eta)}$, shown in Fig.~\ref{figure1}~(c), 
gives an anomalous dimension of $\eta=0.11(2)$.

\begin{figure}[tbp]
	\centering
	\includegraphics[width=0.7\linewidth,clip]{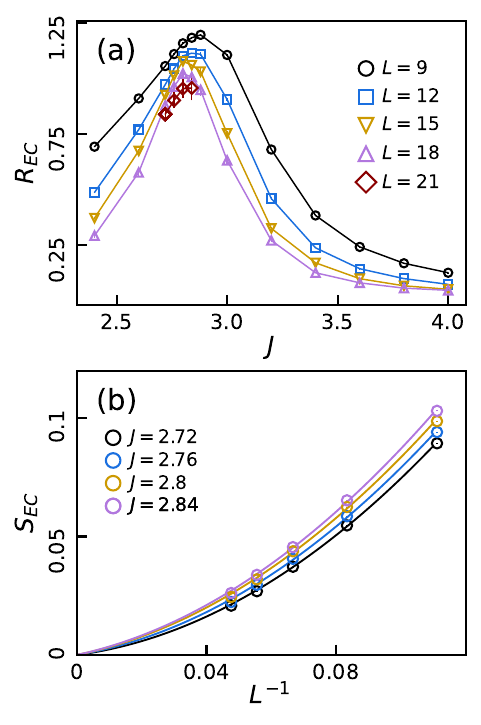}
	\vskip-3mm
	\caption{{\bf Exclusion of the exciton order.} \textbf{(a)} The correlation-length ratio for the exciton order $R_{\text{EC}}$ as a function of $J$ is shown for different system sizes. The lack of crossings in these curves suggests that no phase transition to long-range EC order occurs. \textbf{(b)} The extrapolated results of the exciton structure factor $S_{\text{EC}}$ show that the structure factor tends to zero for $J>J_{\text{c}}$.}
	\label{fig_ec}
\end{figure}

\begin{figure*}[btp]
	\centering
	\includegraphics[width=1.0\linewidth,clip]{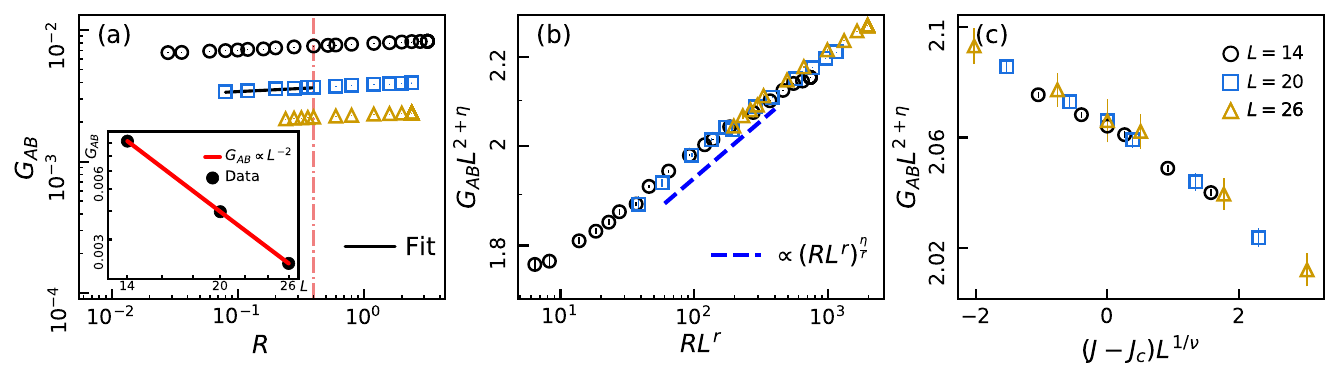}
	\vskip-3mm
	\caption{{\bf Nonequilibrium dynamic scaling of the SMG transition.} \textbf{(a)} Log-log plots of $G_{\text{AB}}$ versus the driving rate $R$ for different $L$ at $J_{\text{c}}=2.584(8)$. The inset in \textbf{(a)} shows $G_{\text{AB}}\propto L^{-2}$ at $R=0.4$, as indicated by the dash-dotted line. For $L=20$, the power-law fit (black solid line) gives $G_{\text{AB}}\propto R^{0.048(6)}$, with an exponent close to $\eta/r=0.053$. \textbf{(b)} In log-log coordinates, the rescaled curves of $G_{\text{AB}}L^{2+\eta}$ versus $RL^r$ collapse onto a single curve, which has a slope close to $\eta/r$ (dashed line). \textbf{(c)} Rescaled curves of $G_{\text{AB}}L^{2+\eta}$ versus $(J-J_{\text{c}})L^{1/\nu}$ for a fixed $RL^{r}=463.92$ and different $L$ collapse onto a single curve.}
	\label{figure2}
\end{figure*}

With the critical properties established, the key open question is the nature of this phase transition and whether it corresponds to the SMG transition. The mean-field theory suggests that the leading alternative to the SMG transition is the EC transition~\cite{houprb2023}. To elucidate the mechanism of the observed phase transition, we examine the correlation-length ratio of the EC order parameter, defined as $R_{\text{EC}}\equiv S_{\text{EC}}(\boldsymbol{q})/S_{\text{EC}}(\boldsymbol{q}+\Delta\boldsymbol{q})-1$, where $S_{\text{EC}}$ is the structure factor (the detailed definition is given in the Supplementary Materials~\cite{SupMat}), $\boldsymbol{q}=0$ and $\Delta\boldsymbol{q}=(2\pi/L,2\pi/L)$. If the observed transition were an EC transition, the dimensionless ratio $R_{\text{EC}}$ would exhibit a crossing at $J_{\text{c}}$ for different system sizes $L$. However, as shown in \Fig{fig_ec}~(a), $R_{\text{EC}}$ displays no such crossing behavior. Rather, $R_{\text{EC}}$ monotonically decreases with increasing $L$, providing compelling evidence for the absence of EC long-range order. This result clearly demonstrates that the observed phase transition is not an EC transition. Additionally, \Fig{fig_ec}~(b) shows the $S_{\text{EC}}$ for $J>J_{\text{c}}$. We find that all extrapolated $S_{\text{EC}}$ for $L\rightarrow\infty$ converge to zero, further suggesting the lack of EC order within this region. In addition, in the Supplementary Materials~\cite{SupMat}, we exclude other possible orders by using a similar procedure. These results demonstrate that the observed quantum phase transition is not driven by a mechanism of spontaneously symmetry breaking and is therefore an SMG transition.

{\it Nonequilibrium criticality}---Then, we turn to the driven critical dynamics of the SMG transition in model (\ref{eq:H1}). The original KZM focus on the driven dynamics in LGW phase transitions characterized by spontaneously symmetry breaking and predicts the scaling behaviors of the driving-induced topological defects, located at the intersections of different broken-symmetry domains, on the driving rate after the quench~\cite{kibble1976,zurek1985}. As a generalization, the FTS theory shows that the general scaling behaviors exist for other quantities in the whole driven process~\cite{zhongprb2005,Gongnjp2010,fengprb2016,huangprb2014}. Although the original KZM and FTS are based on the adiabatic-impulse scenario, which typically requires a gapped initial state to ensure the existence of an adiabatic stage before the system enters the impulse regime near the critical point, it has recently been shown that in symmetry-breaking phase transitions in Dirac systems, the KZM and FTS still hold when the initial state is the gapless DSM phase~\cite{zzarxiv2024,zengFTSKZM2025}. Here, we investigate whether the FTS remains applicable to the SMG transition.

We focus on the dynamic scaling behaviors of the fermion correlation function $G_{\text{AB}}(L)$. In the initial DSM phase, the dimension of $G_{\text{AB}}(L)$ is $[G_{\text{AB}}]=2$ and $G_{\text{AB}}(L)\propto L^{-2}$; while at the critical point, the dimension $[G_{\text{AB}}]=2+\eta$. For driven dynamics to the critical point, when the driving rate $R$ is large, the initial state information should still be preserved at the critical point, making $G_{\text{AB}}(L)\propto L^{-2}$. Therefore, the residual anomalous dimension $\eta$ in $[G_{\text{AB}}]$ should manifest as the dependence on the driving rate $R$. Accordingly, for large $R$, the dynamic scaling relation of $G_{\text{AB}}(L)$ at the critical point $J_{\text{c}}$  should satisfy
 \begin{equation}
 	G_{\text{AB}}(R,L) \propto L^{-2}R^{\frac{\eta}{r}},
 	\label{eq3}
 \end{equation}
 where $ r = z + 1/\nu$ is the dimension of $R$.
In addition, for small $R$, the equilibrium scaling $G_{\text{AB}}(L)\propto L^{-2-\eta}$ must be restored. This requires that $G_{\text{AB}}(R,L)$ should obey the scaling form
 \begin{equation}
	G_{\text{AB}}(R,L) = L^{-2-\eta} \mathcal{F}_1(RL^{r}),
	\label{eq4}
\end{equation}
in which $\mathcal{F}_1(RL^{r})$ should satisfy $\mathcal{F}_1(RL^{r})\propto (RL^{r})^{\eta/r}$ for large $R$.

Fig.~\ref{figure2}~(a) shows the numerical results of the curve of $G_{\text{AB}}(L)$ versus $R$ for different $L$ at $J_{\text{c}}$. Fit with a power function on the curve for $L=20$ shows that $G_{\text{AB}}(R,L)\propto R^{0.048(6)}$, in which the exponent is close to $\eta/r=0.053(10)$ with $\eta=0.11$ and $\nu=0.945$ determined above as inputs. In addition, for a large $R$, the inset shows that $G_{\text{AB}}(L)\propto L^{-2}$. Combining all these results, we thereby establish the validity of Eq.~(\ref{eq3}). Then, by rescaling $G_{\text{AB}}$ and $R$ as $G_{\text{AB}}L^{2+\eta}$ and $RL^r$, respectively, we find that the rescaled curves match with each other, as shown in Fig.~\ref{figure2}~(b), confirming Eq.~(\ref{eq4}). In addition, for large $R$, from Fig.~\ref{figure2}~(b), we find the scaling function $\mathcal{F}_1(RL^{r})$, which is just the rescaled curve, satisfies a power law with the exponent close to $\eta/r$, confirming that $\mathcal{F}_1(RL^{r})\propto R^{\eta/r}$.

To further reveal the dynamic scaling, we now turn to the off-critical-point effects in the whole driven process. For $G_{\text{AB}}$, the FTS analysis predicts that the scaling form should obey
\begin{equation}
	G_{\text{AB}}(R,L,J-J_{\text{c}}) = L^{-2-\eta}\mathcal{F}_2[RL^{r},(J-J_{\text{c}})L^{1/\nu}],
	\label{eqfts}
\end{equation}
To examine Eq.~(\ref{eqfts}), we fix $RL^{r}$ to be a constant and calculate the fermion correlation function for different $(J-J_{\text{c}})$. By  rescaling $G_{\text{AB}}$ and $(J-J_{\text{c}})$ by $L^{2+\eta}$ and $L^{\frac{1}{\nu}}$ respectively, as shown in \Fig{figure2}~(c), we find that the rescaled curves collapse onto each other, thereby confirming Eq.~(\ref{eqfts}). These results are twofold in their significance. First, they successfully generalize the KZM/FTS framework to the SMG phase transition, which lacks a symmetry-breaking order parameter. Second, they provide decisive and independent support to the validity of the critical point and exponents derived from equilibrium analyses.

{\it Discussion}---The conventional KZM relies fundamentally on the phase transition mechanism of spontaneous symmetry breaking. In fact, the driving-induced length scale $\xi_d\propto R^{-1/r}$, a key quantity in the KZM, just quantifies the typical spacing of topological defects at the intersections of distinct broken-symmetry domains. However, although no symmetry breaking is present in the SMG phase transition, the dynamic scaling behaviors shown above demonstrate that $\xi_d$ nonetheless still exists. This poses a key conceptual question: what is the physical meaning of $\xi_d$?

To address this question, we resort to the putative field theory describing the SMG phase transition. In usual symmetry-breaking phase transitions, the order-parameter field provides the fundamental critical fluctuations. In contrast, it was shown that the fundamental building blocks of the field theory describing the SMG phase transition are the fractionalized bosonic and fermionic partons, arising from the decomposition of physical fermions, together with the emergent gauge field~\cite{yzyprx2018,yzyprb2018}. This suggests that $\xi_d$ may characterize the typical fluctuation scales of these fractionalized excitations under external driving. By identifying this scale, we thereby develop a nonequilibrium approach for probing the hidden characteristic length associated with fractionalization.

{\it Summary}---In this paper, we utilize QMC to comprehensively investigate the properties of the bilayer model~(\ref{eq:H1}) on the honeycomb lattice. Our results provide unambiguous evidence for the existence of an SMG phase transition in model~(\ref{eq:H1}), ruling out the possibility of a phase transition associated with any symmetry-breaking phase. Additionally, we determine the critical exponents $\nu=0.945(5)$ and the anomalous dimension $\eta = 0.11(2)$. Notably, these values deviate significantly from those obtained via VMC methods, highlighting the need for further investigation within a field-theoretic framework.

Moreover, by studying the driven critical dynamics of the SMG phase transition, our work firmly establishes that FTS is not tied to the symmetry-breaking mechanism; instead, here it provides a dynamic approach to quantify the fluctuation scales of fractionalized excitations, constituting a notable generalization of the traditional KZM. Consequently, our findings represent a fundamental advance in the nonequilibrium dynamics of phase transitions beyond the LGW framework.

{\it Note added}---Upon completing this work, we became
aware of interesting related papers by He, You, and Xu that focus on the equilibrium criticality in the same model~\cite{he2026,he2026Continuous}. Consistent equilibrium results are obtained therein.

{\it Acknowledgments}---We would like to thank Yuan-Yao He and Wei-Xuan Chang for helpful discussions. Zhi-Xuan Li and S. Yin are supported by the National Natural Science Foundation of China (Grants No. 12222515 and No. 12075324). Y. K. Yu and Zi-Xiang Li are supported by Beijing Natural Science Foundation (F251001), the National Natural Science Foundation of China (No.12347107 and No.12474146) and the New Cornerstone Investigator Program. S. Yin is also supported by Research Center for Magnetoelectric Physics of Guangdong Province (Grant No. 2024B0303390001), the Guangdong Provincial Key Laboratory of Magnetoelectric Physics and Devices (Grant No. 2022B1212010008), and the Science and Technology Projects in Guangzhou City (Grant No. 2025A04J5408).

\bibliographystyle{apsrev4-1}
\bibliography{ref}

\onecolumngrid
\newpage
\widetext
\thispagestyle{empty}

\setcounter{equation}{0}
\setcounter{figure}{0}
\setcounter{table}{0}
\renewcommand{\theequation}{S\arabic{equation}}
\renewcommand{\thefigure}{S\arabic{figure}}
\renewcommand{\thetable}{S\arabic{table}}

\pdfbookmark[0]{Supplementary Materials}{SM}
\begin{center}
    \vspace{3em}
    {\Large\textbf{Supplementary Materials for}}\\
    \vspace{1em}
    {\large\textbf{Symmetric Mass Generation Transition and its Nonequilibrium Critical Dynamics in a Bilayer Honeycomb Lattice Model}}\\
    \vspace{1.5em}
\end{center}

First, we provide the details of the determinant quantum Monte Carlo~(DQMC) method employed to calculate the observables in the ground state and those for the driven dynamics~(Supplementary Note 1). Second, we present the structure factors and correlation-length ratios for charge density wave~(CDW), spin density wave~(SDW), and superconductivity~(SC) orders. These results demonstrate the absence of phase transitions to CDW, SDW, and SC order~(Supplementary Note 2). Finally, we also show examples of how the single-particle gap is determined~(Supplementary Note 3).

\section*{Supplementary Note 1: Determinant quantum Monte Carlo}
\label{sec:S1}
In this paper, we use DQMC to study the ground state properties and nonequilibrium processes in interacting fermion systems. The ground state expectation value of an observable $\hat{O}$ can be obtained by projecting a trial wave function $|\Psi_T\rangle$ along the imaginary time axis. When the projection time $\Theta$ is long enough, the trial initial state can be projected to the ground state, $|\Psi_{0}\rangle\propto\lim\limits_{\Theta\rightarrow\infty}e^{-\Theta\hat{H}}|\Psi_{T}\rangle$. The expectation value of the ground state observables can be expressed as follows:

\begin{equation}
	\langle\hat{O}\rangle = \frac{\langle\Psi_{0}|\hat{O}|\Psi_{0}\rangle}{\langle\Psi_{0}|\Psi_{0}\rangle}=\lim\limits_{\Theta\rightarrow\infty}\frac{\langle\Psi_{T}|e^{-\Theta\hat{H}}\hat{O}e^{-\Theta\hat{H}}|\Psi_{T}\rangle}{\langle\Psi_{T}|e^{-2\Theta\hat{H}}|\Psi_{T}\rangle} .
	\label{S1}
\end{equation}
In this work, we set $2\Theta = 2L+6$ to guarantee the convergence to the ground state.

To accurately implement the projection, the Trotter decomposition is used to discretize the imaginary time. The form of the Trotter decomposition is:
\begin{equation}
	e^{-\Theta\hat{H}}=(e^{-\Delta\tau\hat{H}_{I}}e^{-\Delta\tau\hat{H}_{t}})^{M} ,
	\label{S2}
\end{equation}
where $\Delta\tau=\frac{\Theta}{M}$ is set to $0.025$, which is small enough to guarantee numerical convergence.
 
In addition, the Hubbard-Stratonovich (HS) transformation is required to decouple the fermionic interaction. To apply the HS transformation, the interaction term of the model is written as follows:
 \begin{equation}
\hat{H}_{I}=J\sum_{i}\boldsymbol{S}_{i,1}\cdot\boldsymbol{S}_{i,2}=\sum_{i}\frac{J}{4}[(S^{z}_{i1}+S^{z}_{i2})^{2}+(S^{x}_{i1}+S^{x}_{i2})^{2}+(S^{y}_{i1}+S^{y}_{i2})^{2}-(S^{z}_{i1}-S^{z}_{i2})^{2}-(S^{x}_{i1}-S^{x}_{i2})^{2}-(S^{y}_{i1}-S^{y}_{i2})^{2}] .
 	\label{S3}
 \end{equation}

In the simulation, for a single term in Eq.~(\ref{S3}), we use the discrete HS transformation~\cite{Assaad2008}:
  \begin{equation}
 	e^{\Delta\tau WO^{2}} = \sum_{l=\pm 1,\pm 2} \gamma(l)e^{\sqrt{\Delta \tau W}\eta(l)O}+\mathcal{O}(\Delta\tau^{4}) ,
 	\label{S4}
 \end{equation}
 where $W = J/4$ and $O$ denotes the quadratic term in Eq.~(\ref{S3}), $\gamma(l)$ and $\eta(l)$ have the following values: $\eta(\pm 1) = \pm\sqrt{2(3-\sqrt{6})}$, $\eta(\pm 2) = \pm\sqrt{2(3+\sqrt{6})}$, $\gamma(\pm 1)=1+\sqrt{6}/3$, $\gamma(\pm 2)=1-\sqrt{6}/3$.

After the HS transformation, the interaction term is expressed as:
   \begin{equation}
 	e^{-\Delta\tau\hat{H}_{I}} = \prod_{\sigma_{1}=x,y,z}\left[\sum_{l_{1}^{\sigma_{1}}=\pm 1,\pm 2}\gamma(l_{1}^{\sigma_{1}})e^{\hat{h}_{1}^{\sigma_{1}}}\right] \times \prod_{\sigma_{2}=x,y,z}\left[\sum_{l_{2}^{\sigma_{2}}=\pm 1,\pm 2}\gamma(l_{2}^{\sigma_{2}})e^{\hat{h}_{2}^{\sigma_{2}}}\right],
 	\label{S5}
 \end{equation}
 where $\hat{h}_{1}^{\sigma_{1}}=\sqrt{\frac{\Delta\tau J}{4\eta}}(l_{1}^{\sigma_{1}})(S_{i1}^{\sigma_{1}}-S_{i2}^{\sigma_{1}})$,$\hat{h}_2^{\sigma_{2}}=i\sqrt{\frac{\Delta\tau J}{4\eta}}(l_{2}^{\sigma_{2}})(S_{i1}^{\sigma_{2}}+S_{i2}^{\sigma_{2}})$. Then, using DQMC, we can calculate the physical observables.
 
For the driven dynamics, we need to prepare an appropriate initial state $|\Psi_0\rangle$ and then linearly vary the interaction strength $J$ with imaginary time. In this work, we prepare the DSM initial state with $J_0=0$ and drive the system to cross the critical point by changing $J$ as $J=J_0+R\tau$. The expectation value of the observable $\hat{O}$ is:
  \begin{align}
 	 \langle \hat{O}(\tau) \rangle &= \frac{\langle \Psi(\tau)|\hat{O}|\Psi(\tau) \rangle}{\langle \Psi(\tau)|\Psi(\tau) \rangle} \\
 	 			   &=\frac{\langle \Psi_0 | e^{-\Theta\hat{H}}U^{\dagger}(\tau,0)\hat{O}U(\tau,0)e^{-\Theta\hat{H}}|\Psi_0\rangle}{\langle \Psi(\tau)|\Psi(\tau)\rangle},
 	\label{S100}
 \end{align}
where $|\Psi(\tau)\rangle = U(\tau,0)|\Psi_0\rangle$, with $U(\tau,0) = \mathcal{T}e^{-\int_{0}^{\tau}d\tau'H(\tau')}$ and $\mathcal{T}$ represents the time-ordering operator.

\section*{Supplementary Note 2: The results of structure factors and correlation-length ratios}
\label{sec:S2}

  \begin{figure}[b]
 	\centering
 	\includegraphics[width=\linewidth,clip]{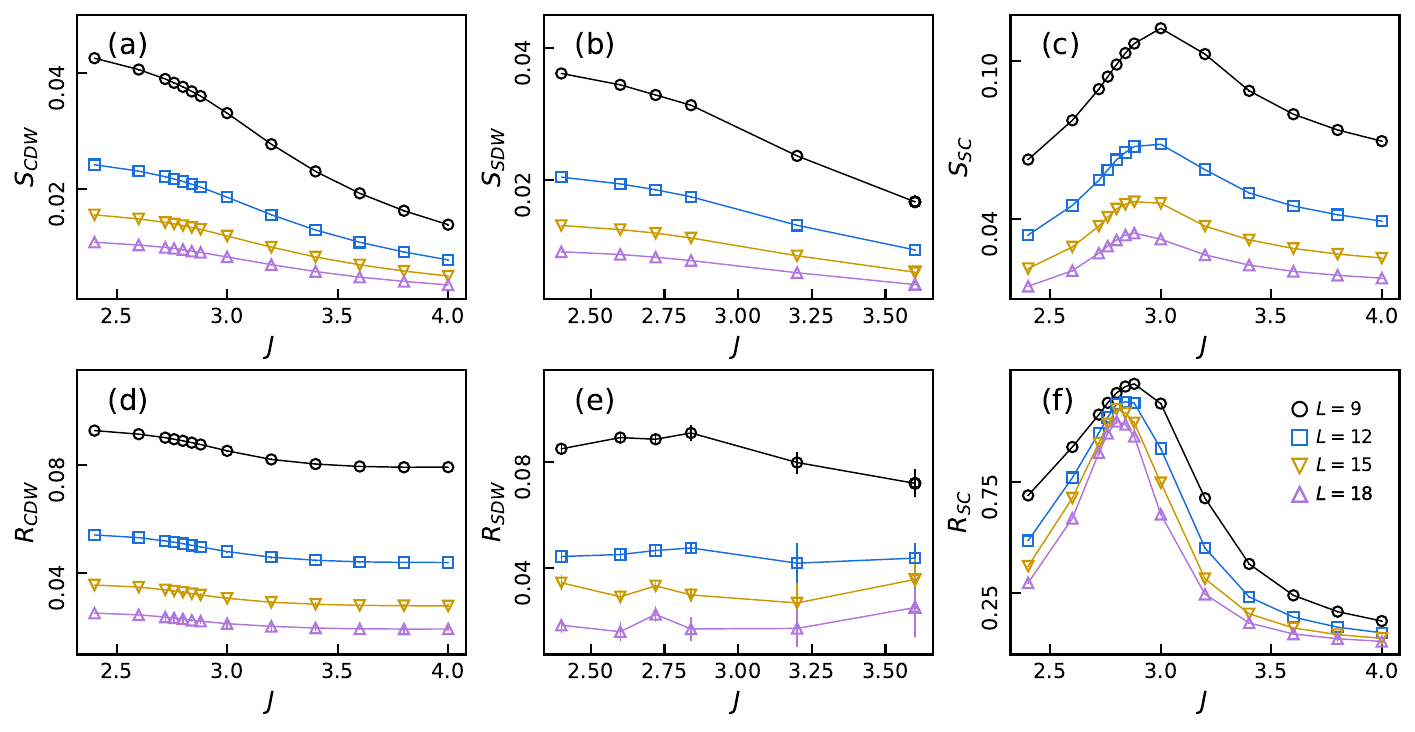}
 	\vskip-3mm
 	\caption{\textbf{Exclusion of CDW, SDW, and SC orders.} The results of the structure factors \textbf{(a)}-\textbf{(c)} and correlation-length ratios \textbf{(d)}-\textbf{(f)} for CDW, SDW and SC as a function of $J$ is shown for different system sizes. The lack of crossings in the curves of the correlation-length ratio suggests that no phase transition to long-range CDW, SDW, and SC order occurs.}
 	\label{figureS1}
 \end{figure}  
 
  \begin{figure}[t]
 	\centering
 	\includegraphics[width=\linewidth,clip]{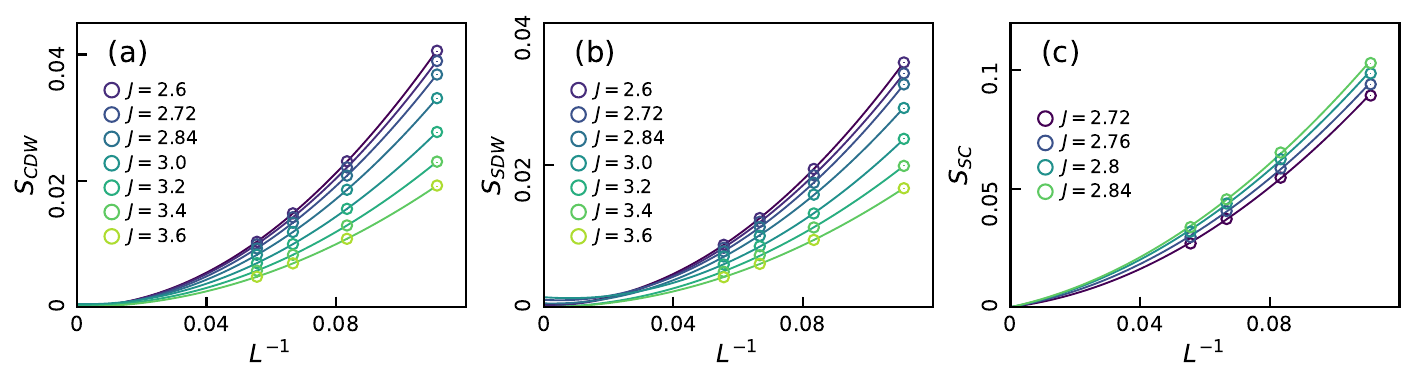}
 	\vskip-3mm
 	\caption{\textbf{The extrapolated results of the structure factors for CDW, SDW, and SC.} The structure factors for $J>J_{\text{c}}$ are extrapolated as a function of $L^{-1}$, and the extrapolated values tend to zero within the error bar in the thermodynamic limit. }
 	\label{figureS2}
 \end{figure}

In the main text, we present the results for the correlation-length ratio of the EC order, based on the corresponding structure factor defined as
$S_{\text{EC}} \equiv L^{-2d} \sum_{i,j} (-)^{i+j} \left\langle 
\left(c^{\dagger}_{i\uparrow 1}c_{i\uparrow 2} + c^{\dagger}_{i\downarrow 1}c_{i\downarrow 2}\right)
\left(c^{\dagger}_{j\uparrow 1}c_{j\uparrow 2} + c^{\dagger}_{j\downarrow 1}c_{j\downarrow 2}\right)
\right\rangle
e^{\text{i}\boldsymbol{q}\cdot(\boldsymbol{r}_i - \boldsymbol{r}_j)}$,
where $i~(j)$ is even (odd) for sublattice A (B) and $\boldsymbol{q} = 0$. Here, we supplement the results for the structure factors of other potential long-range orders and the corresponding correlation-length ratios. The structure factor is defined as follows:
 \begin{equation}
 	S(\boldsymbol{q})\equiv \frac{1}{L^{2d}}\sum_{i,j}\langle A_{i}A_{j}\rangle e^{\text{i}\boldsymbol{q}\cdot(\boldsymbol{r}_i-\boldsymbol{r}_j)} ,
 	\label{S6}
 \end{equation}
 where $A_{i} (A_{j})$ denotes the field operator and $\boldsymbol{q}$ is the momentum. We calculate the structure factors for CDW, SDW and SC orders at $\boldsymbol{q} = 0$ as shown in Fig.~\ref{figureS1}~(a)-(c). Their field operators are defined as~\cite{wxcarxiv2023}:
 \begin{equation}
 	A_{\mathbf{CDW}}(i) \equiv (-)^{i}\sum_{l}(c^{\dagger}_{i\uparrow l}c_{i\uparrow l} + c^{\dagger}_{i\downarrow l}c_{i\downarrow l}) ,
 	\label{S7}
 \end{equation}
 \begin{equation}
	A_{\mathbf{SDW}}(i) \equiv (-)^{i}\sum_{l}(c^{\dagger}_{i\uparrow l}c_{i\uparrow l} - c^{\dagger}_{i\downarrow l}c_{i\downarrow l}) ,
	\label{S8}
 \end{equation}
 \begin{equation}
	A_{\mathbf{SC}}(i) \equiv c^{\dagger}_{i\uparrow 1}c^{\dagger}_{i\downarrow 2} - c^{\dagger}_{i\downarrow 1}c^{\dagger}_{i \uparrow 2} ,
	\label{S9}
 \end{equation}
where $l=1,2$ represents the index of the layer and $i$ is even (odd) for sublattice A (B).
 
In addition, in Fig.~\ref{figureS1}~(d)-(f) we show the correlation-length ratio of different system sizes, which is defined as:
  \begin{equation}
 	R(L)\equiv \frac{S(\boldsymbol{q},L)}{S(\boldsymbol{q}+\Delta\boldsymbol{q},L)}-1 ,
 	\label{S11}
 \end{equation}
where $\Delta\boldsymbol{q} = (\frac{2\pi}{L},\frac{2\pi}{L})$. The correlation-length ratios for CDW, SDW, and SC orders as a function of $J$ are shown for different system sizes in Fig.~\ref{figureS1}~(d)-(f). The lack of crossings in these curves suggests that no phase transition to long-range CDW, SDW, and SC order occurs.

To further investigate whether spontaneous symmetry breaking occurs, we compute the structure factors of CDW, SDW, and SC orders for different system sizes in the region of $J>J_{\text{c}}$, and the extrapolations of these structure factors to the thermodynamic limit are shown in Fig.~\ref{figureS2}~(a)-(c). All extrapolated values tend to zero within the error bar. These results suggest the absence of long-range order corresponding to CDW, SDW, and SC during the phase transition.

\section*{Supplementary Note 3: Single-particle gap from imaginary-time Green's function}
\label{sec:S3}

   \begin{figure}[hb]
 	\centering
 	\includegraphics[width=0.7\linewidth,clip]{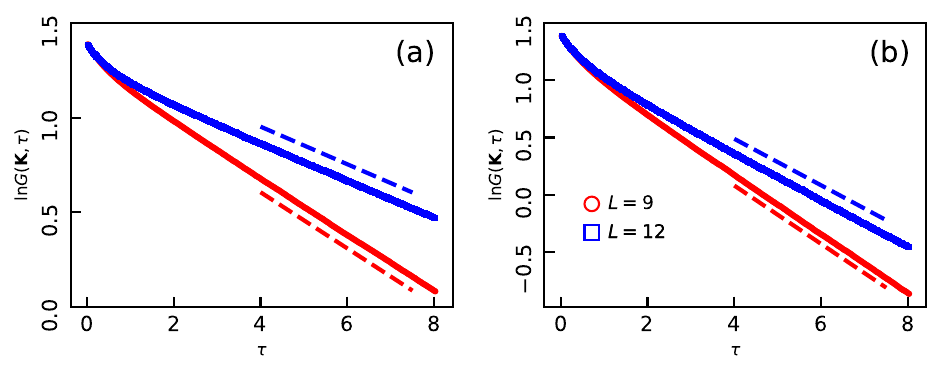}
 	\vskip-3mm
 	\caption{\textbf{Determination of single-particle gap.} $\ln G(\boldsymbol{K},\tau)$ as a function of imaginary time $\tau$. 
\textbf{(a)} For $J = 2.4$ in the DSM phase, the single-particle gaps are determined to be $\Delta_{\text{sp}} = 0.148(1)$ for $L = 9$ and $0.099(1)$ for $L = 12$. The red dashed line and blue dashed line represent the fitted lines for $L = 9$ and $L = 12$, respectively. 
\textbf{(b)} For $J = 2.68$ in the SMG phase, the single-particle gaps are determined to be $\Delta_{\text{sp}} = 0.254(2)$ for $L = 9$ and $0.202(1)$ for $L = 12$.  The red dashed line and blue dashed line represent the fitted lines for $L = 9$ and $L = 12$, respectively.}
 	\label{figureS3}
 \end{figure}  

As discussed in the main text, the single-particle gap $\Delta_{\text{sp}}$ can be obtained from the imaginary-time Green's function $G(\boldsymbol{K},\tau)$. For sufficiently large $\tau$, $G(\boldsymbol{K},\tau)$ satisfies $G(\boldsymbol{K},\tau)\propto e^{-\Delta_{\text{sp}}(\boldsymbol{K})\tau}$. 

Fig.~\ref{figureS3}~(a) and~(b) show $\ln G(\boldsymbol{K},\tau)$ as a function of imaginary time $\tau$ for different system sizes at $J=2.4$ in the DSM phase and $J=2.68$ in the SMG phase, respectively. By performing a linear fit to $\ln G(\boldsymbol{K},\tau)$, the single-particle gaps $\Delta_{\text{sp}}$ are determined as $0.148(1)$ and $0.099(1)$ for $L = 9$ and $L = 12$ at $J = 2.4$, respectively. Following the same procedure, at $J=2.68$, the corresponding values of $\Delta_{\text{sp}}$ for $L=9$ and $L=12$ are $0.254(2)$ and $0.202(1)$, respectively.

\end{document}